\documentclass[prl,nobalancelastpage,twocolumn,superscriptaddress,showpacs,nofootinbib]{revtex4-1}
\usepackage{xcolor,amsthm,amsmath,amsfonts,graphicx,bm,amssymb,subfigure}
\usepackage{epstopdf}
\usepackage{ulem}
\usepackage{soul}

\newcommand{\MyTitle}{First-Order Topological Quantum Phase Transition in a Strongly Correlated Ladder  }

\date{\today}

\begin{document}

\title{\MyTitle}

\author{Simone Barbarino}
\affiliation{Institute of Theoretical Physics, Technische Universit\"at Dresden, 01062 Dresden, Germany}

\author{Giorgio Sangiovanni}
\affiliation{Institute of Theoretical Physics and Astrophysics, Universit\"at W\"urzburg, 97074 W\"urzburg, Germany}

\author{Jan Carl Budich}
\affiliation{Institute of Theoretical Physics, Technische Universit\"at Dresden, 01062 Dresden, Germany}

\begin{abstract}
We report on the discovery of a quantum tri-critical point (QTP) separating a line of first-order topological quantum phase transitions from a continuous transition regime in a strongly correlated one-dimensional lattice system. Specifically, we study a fermionic four-leg ladder supporting a symmetry-protected topological insulator phase in the presence of on-site interaction, which is driven towards a trivial gapped phase by a nearest-neighbor interaction. Based on DMRG simulations, we show that, as a function of the interaction strength, the phase transition between the topological and the trivial phase switches from being continuous to exhibiting a first-order character. Remarkably, the QTP as well as the first-order character of the topological transition in the strongly correlated regime are found to clearly manifest in simple local observables.
\end{abstract}

\maketitle

{\it Introduction.-}
Taking into account topological properties has fundamentally broadened the notion of phase transitions~\cite{Vojta_2003,Sachdev_2011} in at least two directions. First, the celebrated work by Berezinskii, Kosterlitz, and Thouless has shown that certain thermal phase transitions are driven by the proliferation of topological excitations with increasing temperature~\cite{Berezinskii_1971,Berezinskii_1972,Kosterlitz_1973}, and are thus qualitatively distinct from conventional transitions characterized by local order parameters~\cite{Landau_1937}. Second, with the advent of topological phases of quantum matter that cannot be adiabatically connected to conventional materials~\cite{Hasan_2010,Qi_2011}, the concept of topological quantum phase transitions (TQPTs) describing the transition between different topological phases hallmarked by the change of a topological invariant~\cite{Zak_1982,Berry_1984} at zero temperature has emerged.

For non-interacting fermions, TQPTs such as the transition between a topological insulator and a trivial gapped phase are well known to be {\it continuous} in the sense that they are accompanied by the closing of the band gap as long as the symmetries of the system are maintained~\cite{Rombouts_2010,Wen_2017,Cats_2018}. The purpose of this work is to investigate at a fully microscopic level how this paradigm is altered in the presence of strong correlations between the particles~\cite{Hohenadler_2013,Rachel_2018}. To this end, we study the stability of a symmetry protected topological insulator (TI) phase against on-site and nearest-neighbor interactions in a one-dimensional (1D) lattice model. While all TQPTs in the presence of on-site interactions only are found to be continuous, switching on nearest neighbor interaction we observe a quantum tricritical point (QTP) from which the transition between the TI phase and a trivial gapped phase is found to be of first-order.

\begin{figure}
	\begin{center}
  	\includegraphics[width=\columnwidth]{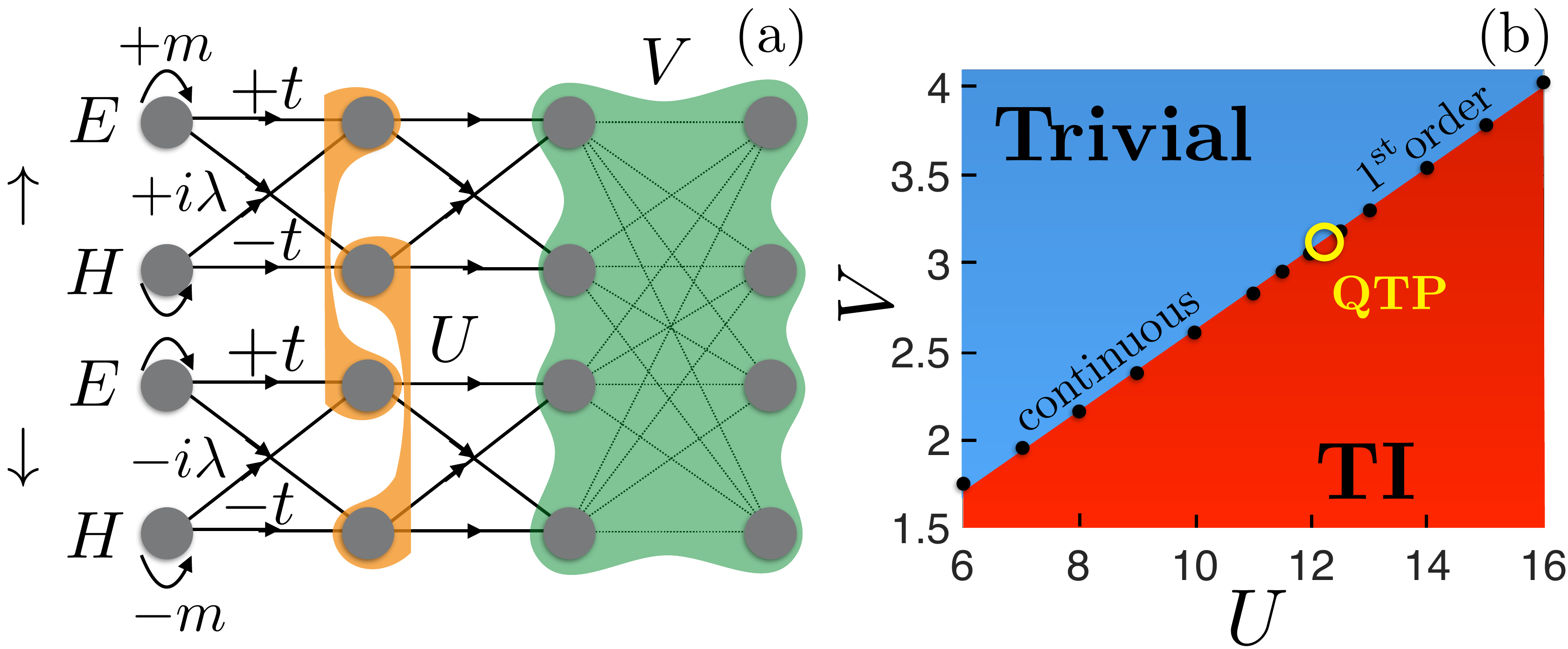}
	\end{center}
	\caption{(a) Illustration of model (see Eqs. (\ref{model_nonint}-\ref{interaction_term})): A 1D chain of spinful fermions with four internal degrees of freedom shown as a four-leg ladder, where particles are allowed to hop horizontally and diagonally (solid lines), subject to on-site interactions of strength $U$ (double lines) and nearest neighbor interactions of strength $V$ (dashed lines). 
	(b) Phase diagram at half filling as a function of $U$ and $V$ (in units of $t$) at parameters $\lambda=t$ and $m=0$, obtained from DMRG data (black dots).  
	The yellow circle indicates the quantum tricritical point (QTP) at which the transition switches from second to first order.
	}
	\label{model}
\end{figure}

Previous work on TQPTs in correlated 2D and 3D TIs has predicted the occurrence of first-order TQPTs with on-site interactions~\cite{Budich_12,Budich_13,Amaricci_2015,Amaricci_2016,Roy_2016,Imriska_2016}, using 
variational cluster approaches and single-site dynamical mean field theory, which is however approximate except in the limit of infinite coordination number~\cite{Georges96}. Our present exact numerical results employing a density matrix renormalization group~\cite{White92,Schollwock_05} (DMRG) approach show that there is no direct analog of this phenomenon in 1D. Instead, the presence of nearest neighbor interaction destabilizing the system towards the formation of a density-wave is crucial for the QTP and the first order topological transition line reported here.

Below, we consider a fermionic four-leg ladder (see Fig.~\ref{model}(a) for an illustration), supporting, in the absence of interactions, a symmetry-protected TI phase in symmetry class DIII of the Altland-Zirnbauer classification~\cite{Altland97,Ludwig15}. From the combination of local correlation functions as well as the bipartite entanglement spectrum~\cite{Fidkowski10,Pollmann10,Berg_11}, we map out the topological phase diagram of the ladder system subject to the interplay of strong repulsive Hubbard interactions and nearest-neighbor interactions. We present a detailed analysis of the nature of the quantum phase transition between the TI phase and a trivial density-wave ordered phase. The aforementioned QTP and the first-order character of the transition at sufficiently strong correlations is identified by a sharp cusp in the ground state energy as well as by discontinuities in simple local observables such as local spin fluctuations.

{\it Microscopic Hamiltonian.-} 
We consider a 1D chain of spinful fermions endowed with four internal states labeled by two independent indices
$\eta$ and $\sigma$ corresponding to an orbital $\eta=E \, (+1), \, H \,(-1)$ and a spin $\sigma=\uparrow \, (+1), \, \downarrow \,(-1)$ degree of freedom, 
respectively. Fermions are annihilated (created) by the operators  $\hat c^\dagger_{j,\eta,\sigma}$ ($\hat c_{j,\eta,\sigma}$) where $j=1,\dots, L$ labels the lattice sites. 
The  Hamiltonian $\hat H=\hat H_0 + \hat H_{\rm int}$ consists of a single-particle contribution 
\begin{align}
& \hat H_0 = \sum_{j=1}^L \sum_{\eta=\pm 1, \sigma=\pm 1}  \left[ \left(t\, \eta \, \hat c^\dagger_{j+1,\eta, \sigma} \hat c_{j,\eta, \sigma}   + \mathrm{H.c.} \right) +\right.\nonumber\\
&\left.   + \left(i  \, \sigma \lambda \,  \hat c^\dagger_{j+1,\eta, \sigma} \hat c_{j,-\eta, \sigma}+\mathrm{H.c.}\right) + m \,\eta \, \hat n_{j,\eta, \sigma} \right]  
\label{model_nonint}
\end{align}
with $t$, $\lambda$, and $m$ real coefficients, $\hat n_{j,\eta, \sigma}=\hat c^\dagger_{j,\eta, \sigma}\hat c_{j,\eta, \sigma}$; and an interaction term 
 \begin{align}
\hat H_{\rm int}= U \sum_j \sum_{\eta=\pm 1} \hat n_{j,\eta,\uparrow}\hat n_{j,\eta,\downarrow} + V\sum_j \hat n_j \hat n_{j+1}\,
\label{interaction_term}
\end{align}
with $\hat n_j= \sum_{\eta, \sigma} \hat n_{j,\eta,\sigma}$. Here $U$ is a density-density intra-orbital Hubbard interaction which couples fermions in the same orbital, and
$V$ corresponds to a nearest-neighbor density-density interaction~\cite{foot1}. 
In total, the 1D chain can be seen as a four-leg ladder~\cite{Boada12}  where spinless fermions are allowed to hop horizontally and diagonally, as shown in Fig.~\ref{model}(a).
\newline  
In momentum space, the Hamiltonian $\hat H_0$ is the direct sum of two Kramers partners $h_0(k)$ and $h_0^*(-k)$ due to the presence of time reversal symmetry (TRS), where $h_0(k)= (2  t \cos k +m)\; \tau_z +2 \lambda \sin k\; \tau_x$. Here, the Pauli matrices $\tau_i$ act on the orbital degree of freedom, while the Pauli matrices $\sigma_i$ act on the spin degree of freedom, and  
the Bloch Hamiltonian can be written as 
\begin{equation}
 H_0(k)= (2  t \cos k +m)\; \tau_z \otimes \sigma_0 +2 \lambda \sin k\; \tau_x \otimes \sigma_z\, ,
\label{BHZ}
\end{equation} 
where $\sigma_0$ denotes the  $2 \times 2$ identity matrix. The spectrum of the Hamiltonian~\eqref{BHZ} consists of two spin-degenerate copies of the bands $E_{\pm}=\pm \sqrt{(2t \cos k +m)^2 + 4\lambda^2 \sin^2 k }$.
At filling $\nu=N/L=2$, the Hamiltonian~\eqref{model_nonint} supports a TI  phase for $m<2t$ and a trivial phase  for $m>2t$ which are separated by a continuous phase transition occurring at $m=2t$, where the system becomes semi-metalic. The TI phase is protected by TRS as well as a particle-hole constraint, thus belonging to the symmetry class DIII of the Altland-Zirnbauer classification. Explicitly, $H_0(k)$ is clearly time-reversal symmetric since $U_T H_0^*(k)U_T^\dagger=H_0(-k) $ with the unitary $U_T=i \sigma_y \otimes \tau_0$. Regarding the particle hole constraint, we note that $U_C H_0^*(k)U_C^\dagger=-H_0(-k)$ with $U_C= i \sigma_x 	\otimes \tau_y $. Both the on-site and the nearest-neighbor interaction terms contained in $\hat H_{\rm int}$ do not break the symmetries protecting the non-interacting topological phase. This is easy to see by considering the action of the symmetries directly on the fermionic operators along the lines of Ref.~\cite{Ludwig15}.
\begin{figure}
	\begin{center}
  	\includegraphics[width=\columnwidth]{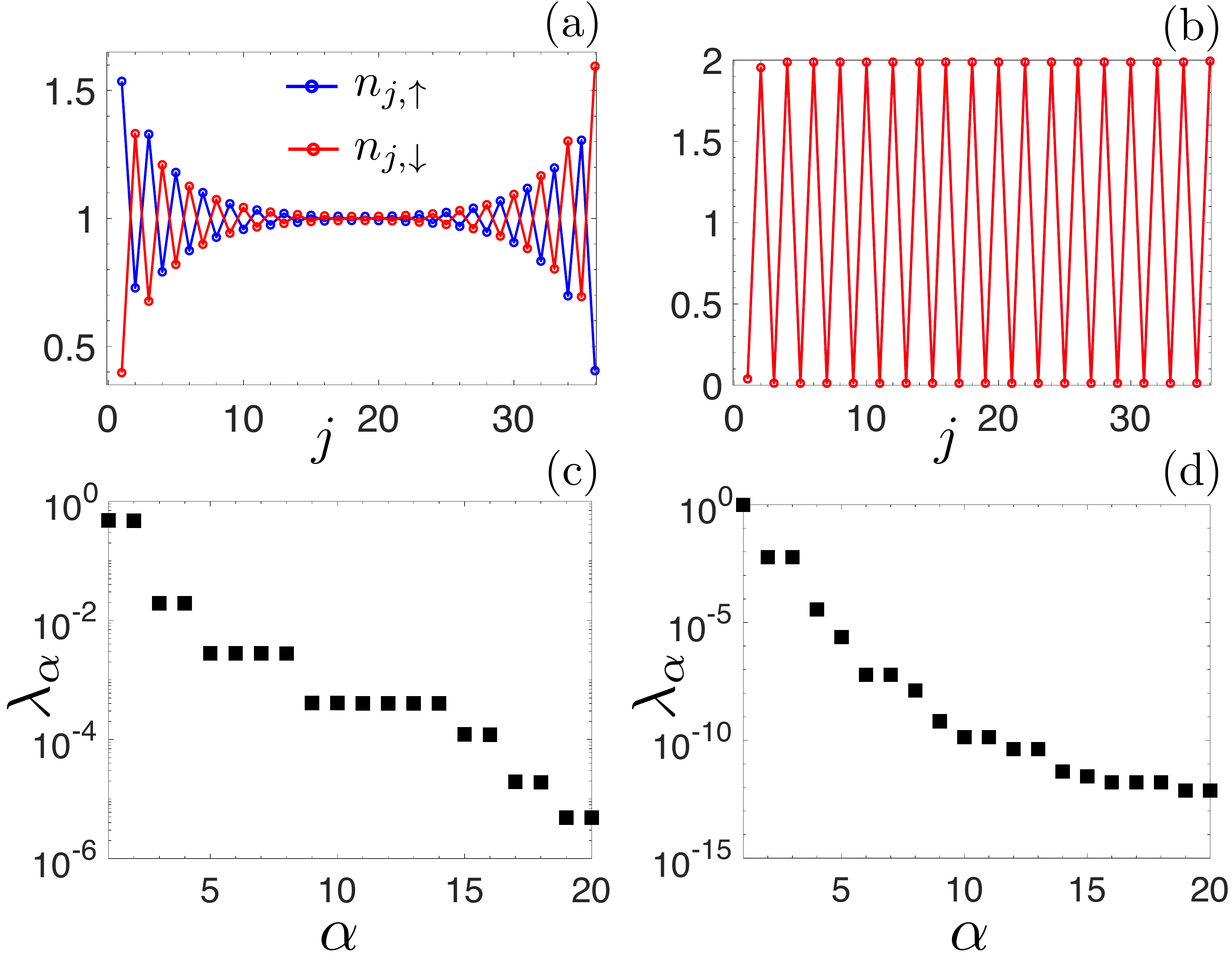}
	\end{center}
	\caption{DMRG data for the ground state expectation value of the density operator $\sum_\eta \hat n_{j,\eta,\sigma}$
	and the highest twenty eigenvalues $\{ \lambda_\alpha \}$ of the entanglement spectrum with $\ell=L/2$; $L=36$ sites. 
	(a,c): $U=16$ and $V=0$; (b,d): $U=16$ and $V=6$. 
		 }
	\label{density_profiles}
\end{figure} 
\begin{figure*}
	\begin{center}
  	\includegraphics[width=2\columnwidth]{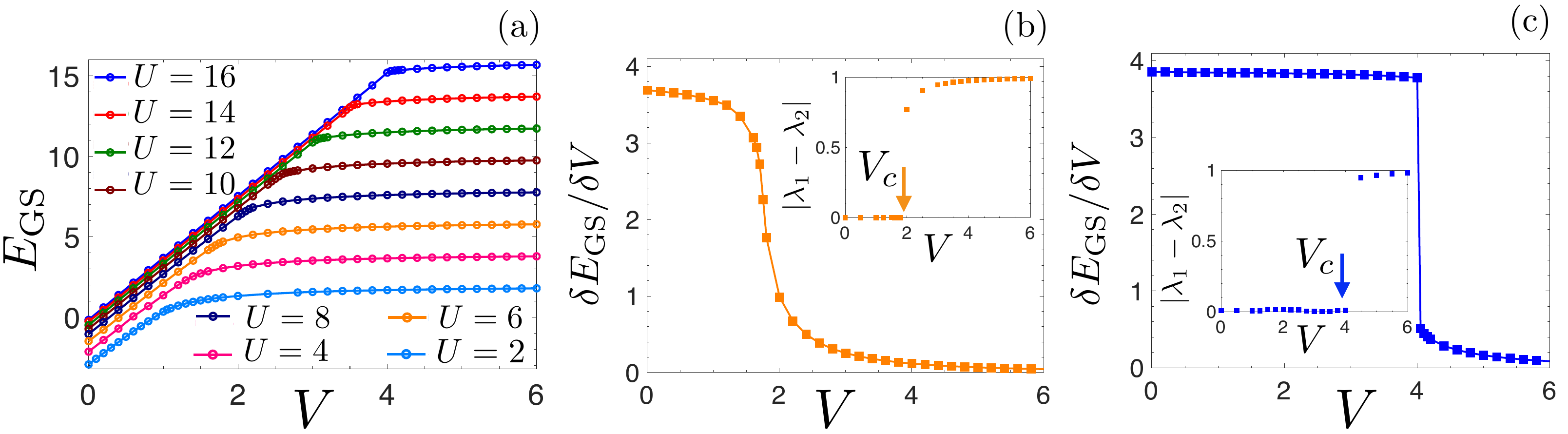}
	\end{center}
	\caption{(a) The ground-state energy $E_{\rm GS}$, see Eq.~\eqref{EGS}, as a function of $V$ for different values of $U$. 
	(b) The derivative for $E_{\rm GS}$ with respect to $V$ for $U=6$ (second-order phase transition);
	(c) the derivative for $E_{\rm GS}$ with respect to $V$ for $U=16$ (first-order phase transition); $\delta V=0.01$.  
	Insets: the difference between the two highest eigenvalues of the entanglement spectrum for $U=6$ and $U=16$ respectively. 
	Here: $L=36$ sites; DMRG data.
	}
	\label{phase_transition}
\end{figure*} 

{\it Numerical analysis.-}  
We proceed with our analysis in two steps. First, we investigate the topological properties of the ground state when the interaction term~\eqref{interaction_term} is switched on assuming that $U>0$ and $V>0$ can be tuned independently. We observe that the topological phase appearing in the non-interacting regime for $m<2t$ is stable for arbitrarily large values of the Hubbard interaction $U$, while the nearest-neighbor interaction $V$ drives the topological phase towards a trivial density-wave ordered state. 
Secondly,  we demonstrate that the phase transition separating the TI phase from the trivial density-wave state obtained by varying $V$ at fixed Hubbard interaction, can be of first order or continuous, depending on the fixed value of the on-site interaction $U$. Resolving this behavior more quantitatively, we determine the position of the QTP where the transition changes its character. 
For simplicity, we consider the fixed band-structure parameter values $\lambda=t$ and $m=0$, and discuss the phase diagram of the model as a function of $U$ and $V$. We carefully checked that our results are only quantitatively, but not qualitatively modified, when $\lambda \neq t$ or $0<m<2t$, and do not rely on finite size effects. For $m>2t$, the trivial non-interacting state is driven into the TI phase by increasing $U$. However, differently from previous work on the 2D~\cite{Amaricci_2015} and 3D~\cite{Amaricci_2016} counterpart of our model, we do not find a QTP and first-order TQPTs on this transition line; in the following, all quantities are expressed in units of $t$.

We start by considering the case of the on-site interaction term $U$ only, i.e. $V=0$. 
The interacting ground state $|\Psi \rangle$ is adiabatically connected to the topological ground state appearing in the non-interacting regime and is stable for arbitrarily large values of the Hubbard interaction $U$. This is shown in Fig.~\ref{density_profiles}(a), where the density profile  $n_{j,\sigma}= \langle \Psi|\sum_\eta \hat n_{j,\eta,\sigma}|\Psi \rangle$ reflects the presence of zero-energy modes exponentially localized at the edges of the chain. In order to further substantiate the topological nature of the ground state, we address the entanglement spectrum which corresponds to the set of the eigenvalues $\{ \lambda_\alpha \}$ of the reduced density matrix $\hat \rho_\ell=\mathrm{Tr}_{\overline{\ell}} \big[ |\Psi\rangle \langle \Psi| \big]$; here $\ell <L$ is a generic subsystem of the entire chain and $\overline{\ell}$ its complement. In our case, a topological (trivial) phase corresponds to a degenerate (non-degenerate) entanglement spectrum. 
As expected, the entanglement spectrum is doubly degenerate, see Fig.~\ref{density_profiles}(c).  
When a sufficiently large nearest-neighbor interaction $V$ is considered, the ground state is a trivial density-wave ordered phase, i.e. a gapped phase characterized by an alternating pattern of empty and fully occupied sites~\cite{foot2}, 
as shown in Fig.~\ref{density_profiles}(b). Accordingly, the entanglement spectrum, is non-degenerate, see Fig.~\ref{density_profiles}(d). 

We now investigate the phase transition occurring between the topological and the trivial phase. 
To this end, in Fig.~\ref{phase_transition}(a), we consider the ground state energy
 \begin{align}
E_{\rm GS} = L^{-1} \langle \Psi|\hat H_0 + \hat H_{\rm int} |\Psi \rangle
\label{EGS}
\end{align}
as a function of $V$ for different values of the on-site interaction $U$ and we show that, depending on the value of the on-site interaction $U$, an either continuous phase transition or a first-order phase transition occurs.  Indeed, the way $E_{\rm GS}$ grows as a function of $V$ strongly depends on the value of the on-site interaction $U$. 
For small values of $U$, $E_{\rm GS}$ increases smoothly, as in standard continuous topological quantum phase transitions.  
On the contrary, when $U$ is sufficiently large, $E_{\rm GS}$ exhibits a sharp kink at the phase transition. 
In Figs.~\ref{phase_transition}(b-c), we have show the derivative of $E_{\rm GS}$ with respect to the nearest-neighbor interaction $V$, i.e.
 \begin{align}
\frac{\delta E_{\rm GS}}{\delta V}(V_m) =\frac{E_{\rm GS}(V+\delta V)-E_{\rm GS}(V)}{\delta V}
\label{derivative_EGS}
\end{align}
with $V_m= 0.5 (V+\delta V)$ and $\delta V=0.01$.
For $U=6$, the continuous character of the phase transition is confirmed by the continuity of the derivative~\eqref{derivative_EGS}. On the contrary, for  $U=16$, 
the derivative exhibits a discontinuity signaling the first-order character of the phase transition. 
The exact value of the critical interaction $V_c$ for which the phase transition occurs can be determined, for both the continuous and the first-order case, by considering the difference $|\lambda_1 -\lambda_2|$ between the two highest eigenvalues of the entanglement spectrum as a function of $V$, see the insets of the Figs.~\ref{phase_transition}(b,c), and taking into account that $|\lambda_1 -\lambda_2|=0$ ($|\lambda_1 -\lambda_2| > 0$) for a topological (trivial)  phase.

\begin{figure}
	\begin{center}
  	\includegraphics[width=\columnwidth]{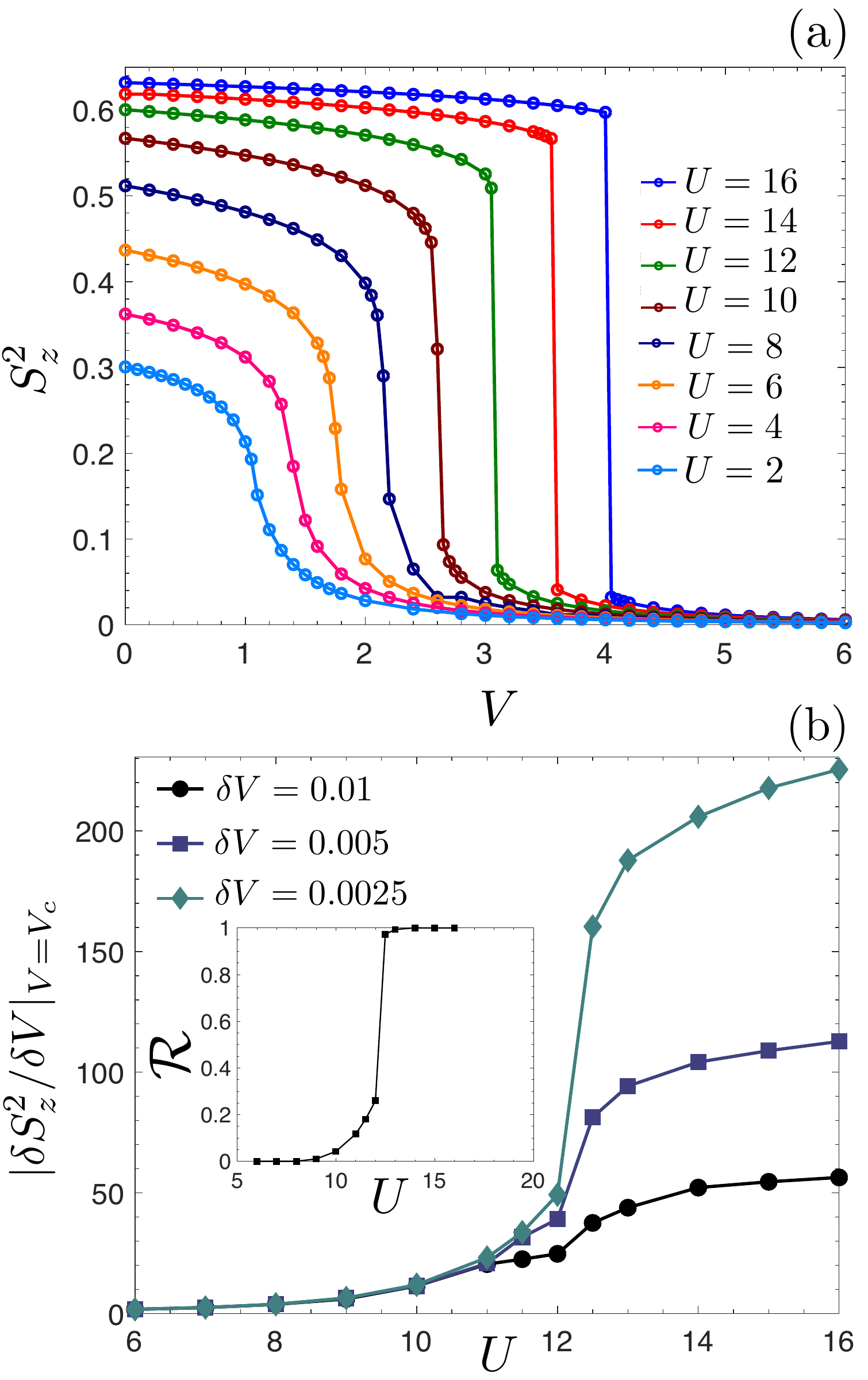}
	\end{center}
	\caption{
	(a) The ground state expectation value of the operator as a function of $V$ for different values of $U$.
	(b) The maximum value of the derivate $|\delta S^2_z / \delta V|_{V=V_c}$ as a function of $U$ for different values of $\delta V$. 
	Inset: the ratio $\mathcal{R}$, see Eq.~\eqref{ratio}, calculated for $\delta V=0.0025$. 
	Here: $L=36$ sites; DMRG data.
	}
	\label{observables}
\end{figure}

Finally, we show that the nature of the phase transition can be diagnosed by studying the behavior of the ground-state expectation value of simple local observables such as the instantaneous 
spin susceptibility
\begin{align}
\left.  S^2_z \equiv \frac 14 \Big \langle \left( \sum_{\eta, \sigma} \sigma\, \hat n_{j,\eta,\sigma}\right)^2 \Big \rangle \right |_{j=\frac L2}   \,.
\label{SZ}
\end{align}
In the non-interacting case, it is easy to show that $S^2_z=0.25$; while, for $U \rightarrow +\infty$ and $V=0$, it saturates to a larger value, i.e. $S^2_z =2/3$; in the trivial phase, on the contrary, $S^2_z=0$.  In Fig.~\ref{observables}(a), we plot $S^2_z$ as a function of $V$ for different values of the on-site interaction $U$ and we observe that, in correspondence of a continuous phase-transition $S^2_z$ is continuous too (its derivative exhibits a peak in correspondence of the phase transition), while, in the case of a first-order phase transition, $S^2_z$ is discontinuous. 

We now use this property to determine the position of the QTP  where the phase transition changes its character. 
To this end, for each value of $U$, we have calculated the maximum value of the derivative $|\delta S^2_z / \delta V|$ for different values of $\delta V$. Similarly to the case of the ground-state energy, the peak of  $|\delta S^2_z / \delta V|$ is observed in correspondence of the phase transition occurring at $V=V_c$.
The derivative $|\delta S^2_z / \delta V|_{V=V_c}$  is shown in Fig.~\ref{observables}(b) as a function of $U$ for different values of $\delta V$. 
If the transition is continuous, the maximum of the derivative converges to a finite value when $\delta V$ is decreased. On  the contrary, 
if the transition is of the first-order, the maximum of the derivative diverges: e.g., when $\delta V$ is decreased by a factor of two, the derivative increases by the same factor.
To further clarify this point, we consider $|\delta S^2_z / \delta V|_{V=V_c}$ as a function of $\delta V$ and we consider the ratio
\begin{align}
\mathcal{R}= \frac{|\delta S^2_z(2\delta V) / \delta V|_{V=V_c}- |\delta S^2_z(\delta V) / \delta V|_{V=V_c}}{|\delta S^2_z(\delta V) / \delta V|_{V=V_c}}
\label{ratio}
\end{align}
which saturates to zero when the transition is continuous and to one when it is of first-order, as shown in the inset of Fig.~\ref{observables}(b) for $\delta V=0.0025$.  The position of the QTP can be then determined taking into account that the phase transition changes its character in correspondence of the interaction $U_c \sim 12.5$ for which  $\mathcal{R}$ exhibits a step-like behavior. The small tail that we observe for $U<U_c$ is expected to disappear if smaller values of $\delta V$ are considered. 

{\it Concluding discussion.-}
In this work, we demonstrated the existence of a quantum tricritical point separating a continuous from a first-order transition line in the quantum phase transition between a symmetry protected TI phase and a trivial gapped phase. Our results represent the first observation of such an intriguing behavior from numerically exact calculations. Representing a genuine qualitative correlation effect in the theory of topological quantum phase transitions, our findings are not only of theoretical interest, but has immediate observable consequences manifesting in discontinuous behavior of simple local observables. 

A particularly promising platform for the direct observation of our predictions on TQPTs is provided by ultracold atoms in optical lattices~\cite{Bloch2008,Goldman2016,Cooper2018}, where the experimental realization 
of multi-leg ladder systems~\cite{Mancini15,Livi16,Kolkowitz17,Kang18,Kang18b} with tunable interactions has already been achieved, and observables such as the local spin fluctuations shown in Fig.~\ref{observables} can be measured with single-site resolution using state of the art quantum gas microscopy methods~\cite{GreinerMicroscope,Weitenberg:2011,Endres,GreinerEntanglement}.

Finally, we note that first-order transitions in 1D lattice models with short-ranged interactions have also been predicted between topologically trivial antiferromagnetic and charge ordered phases~\cite{Nakamura_2000,Sengupta_2002,Ejima_2007,Zhang_2004,Glocke_2007}. Here, by contrast, we report on the observation of first-order quantum phase transitions between two time-reversal symmetric states, one of which represents a symmetry protected TI phase. While our finite size model system exhibits edge magnetic order that exponentially decays into the bulk (see Fig.~\ref{density_profiles}), we carefully analyzed that the first-order character of the transition in our model is a bulk effect that persists in the thermodynamic limit.

{\it Acknowledgements.-} We acknowledge fruitful discussions with M. Dalmonte, M. Vojta, and B. Trauzettel.
We are grateful to D. Rossini for kindly providing us with the DMRG code. J.C.B. acknowledges financial support from the German Research Foundation (DFG) through the Collaborative Research Centre SFB 1143.
G.S. acknowledges support from the DFG through SFB 1170 ToCoTronics (project C07).


\begin{thebibliography}{99}
\bibitem{foot2}
In correspondence of the first site we have put a small pinning potential in order to facilitate the convergency of the DMRG algorithm towards one of the two equivalent configurations of the
density-wave ordered phase.
\end{thebibliography}
\end{document}